\def\astroph{1}

\ifnum\astroph=0
\documentclass[12pt,preprint]{aastex}

\else
\documentclass{emulateapj}
\fi

\usepackage{natbib}
\newcommand{\CAPI}{Histogram showing the distribution of
                horizontal distances a field line travels from the
                photosphere to the lower corona for a typical snapshot
                (full) and a potential extrapolation of the same
                field (dashed)}

\newcommand{\CAPII}{Histogram showing the difference between foot
                point locations for field lines going through the same
                points in the lower corona, for a potential
                extrapolation and a snapshot}

\newcommand{\CAPIII}{Foot point widths in $x$ and $y$ at the photosphere 
                (dots) and in the transition region (crosses) for 
                loops with a circular cross section and a diameter of 
                0.8 Mm at their top. Foot points of bright TRACE loops 
                are encircled}

\newcommand{\CAPVI}{Emulated TRACE 171 (top) and 195 (middle)
                with an emulated magnetogram (bottom), showing three
                loops bright in emission, and 5 points as crosses in
                the emulated TRACE 171 image which are used to create
                DEM curves}

\newcommand{\CAPVII}{Loop 1 with left and right leg in
                Fig. \ref{fig:magneto70} being to the left and right
                respectively, for the  
                energy deposition showing heating (solid),
                radiative cooling (dashed), Spitzer conductivity
                (dash-dot-dot-dot), convective energy deposition
                (long dashes) and emission in the TRACE 171 {\AA}
                filter(dotted) as a function of height}

\newcommand{\CAPVIII}{Gas parameters for loop 1 showing density
                (solid), gas pressure (dashed), 
                temperature (dotted) and the absolute velocity along
                the magnetic field (long dashed). For this loop
                the gas flows down both legs}

\newcommand{\CAPIX}{Similar to Fig. \ref{fig:loop1-171-energy}, here
                for loop 2}

\newcommand{\CAPX}{Similar to Fig. \ref{fig:loop1-171-gas}, here for
                loop 2 with many reversals in gas flow direction}

\newcommand{\CAPXI}{Similar to Fig. \ref{fig:loop1-171-energy}, here
                for loop 3}

\newcommand{\CAPXII}{Similar to Fig. \ref{fig:loop1-171-gas}, here for
                loop 3 with gas flows down both legs from a position just below
                the loop top in the left leg}

\newcommand{\CAPXIII}{Emission in the TRACE 171 filter as function of
  time for loop 1. The dashed line shows the time of \Fig{fig:magneto70}}

\newcommand{\CAPXIV}{DEM curves for the five points along loop 1 shown
  in Fig.\ \ref{fig:magneto70}. The points are numbered from left to right
  with numbers 1 to 5. Curves 2-5 are shifted up by 1-4 orders of
  magnitude respectively}

\newcommand{\CAPXV}{Emission contribution (solid) and temperature
  (dashed) along the projected axis at point 4 of loop 1 shown in
  \Fig{fig:magneto70} (top)}

\ifnum\astroph=0
 
        \newcommand{\FIGI}{
                \begin{figure}[t]
                \figurenum{1}
                \plotone{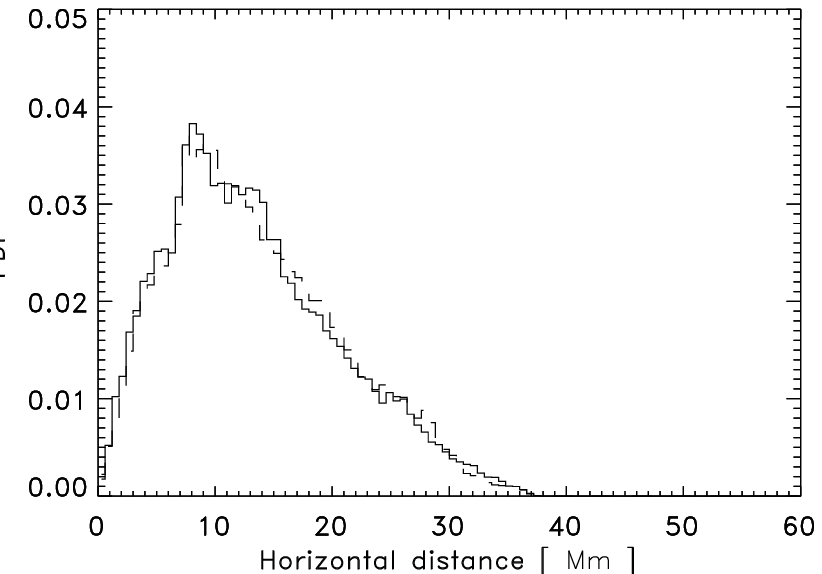}
                \caption{\CAPI}
                \label{fig:connection1}
                \end{figure}
        }
 
        \newcommand{\FIGII}{
                \begin{figure}[t]
                \figurenum{2}
                \plotone{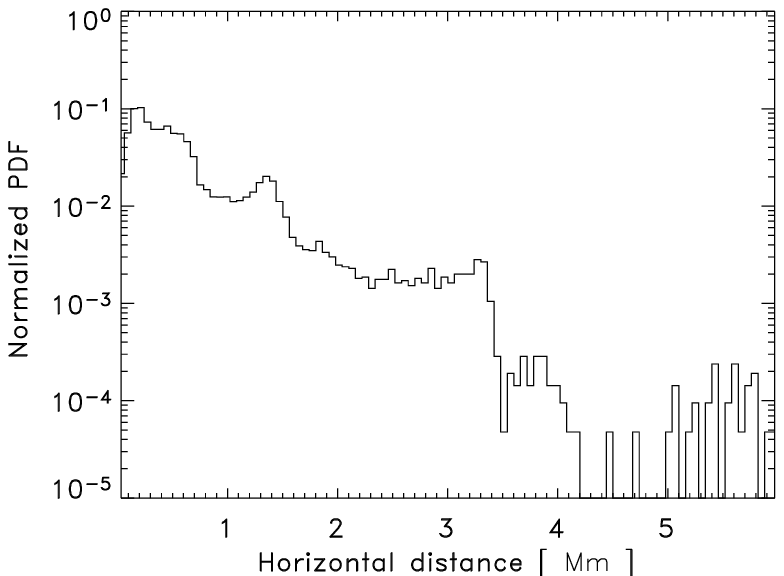}
                \caption{\CAPII}
                \label{fig:connection2}
                \end{figure}
        }

        \newcommand{\FIGIII}{
                \begin{figure}[t]
                \figurenum{3}
                \plotone{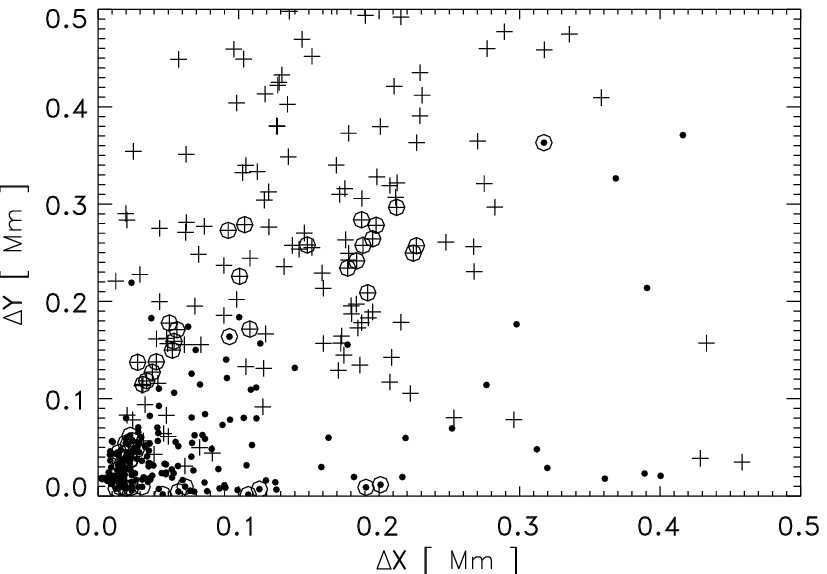}
                \caption{\CAPIII}
                \label{fig:footpoints}
                \end{figure}
        }        

        \newcommand{\FIGIV}{
                \begin{figure}[t]
                \figurenum{4a}
                \plotone{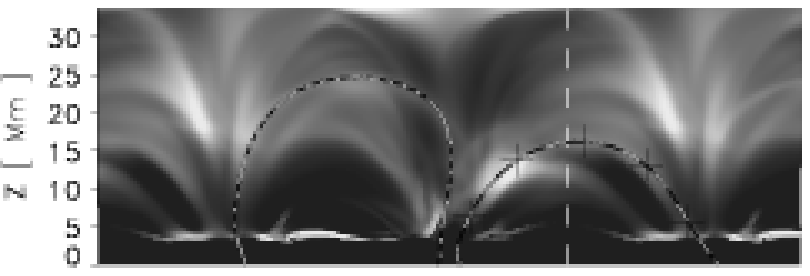}
                \label{fig:loop-tr171}
                \end{figure}
        }
        
        \newcommand{\FIGV}{
                \begin{figure}[t]
                \figurenum{4b}
                \plotone{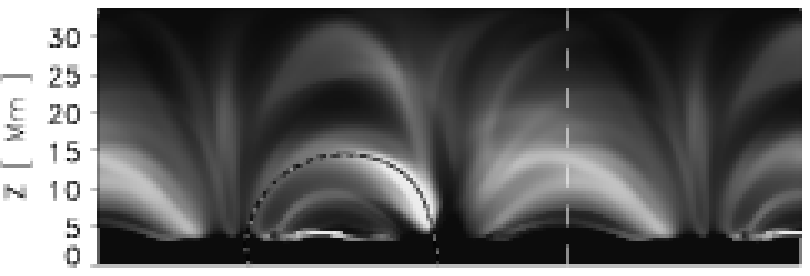}
                \label{fig:loop-tr195}
                \end{figure}
        }

        \newcommand{\FIGVI}{
                \begin{figure}[t]
                \figurenum{4}
                \plotone{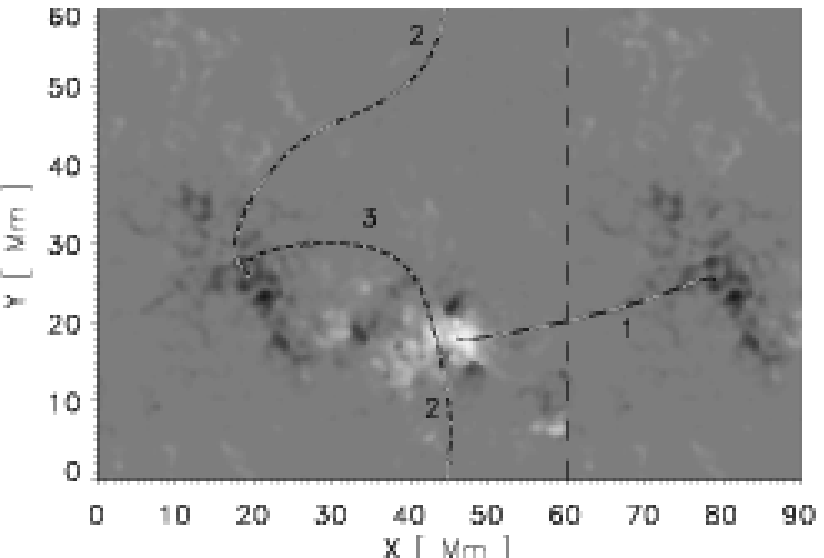}
                \caption{\CAPVI}
                \label{fig:magneto70}
                \end{figure}
        }

        \newcommand{\FIGVII}{
          \begin{figure}[t]
            \figurenum{5}
            \plotone{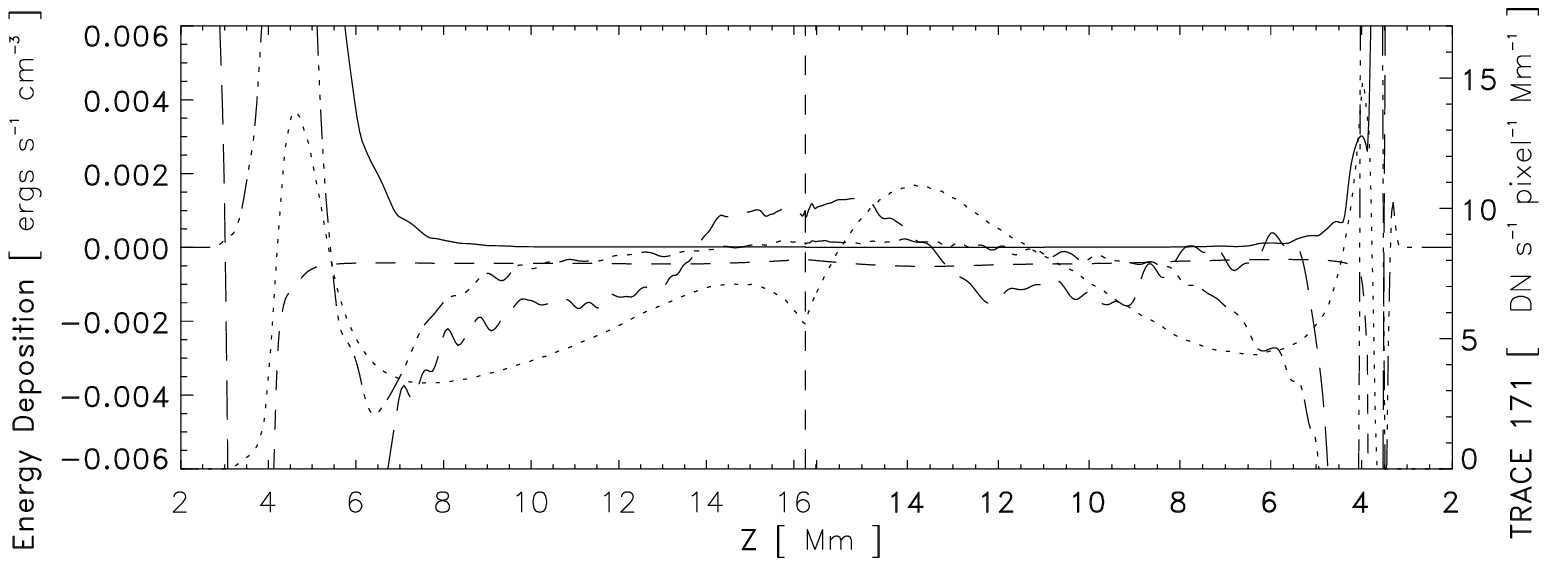}
            \label{fig:loop1-171-energy}
            \caption{\CAPVII}
          \end{figure}
        }

       \newcommand{\FIGVIII}{
         \begin{figure}[t]
           \figurenum{6}
           \plotone{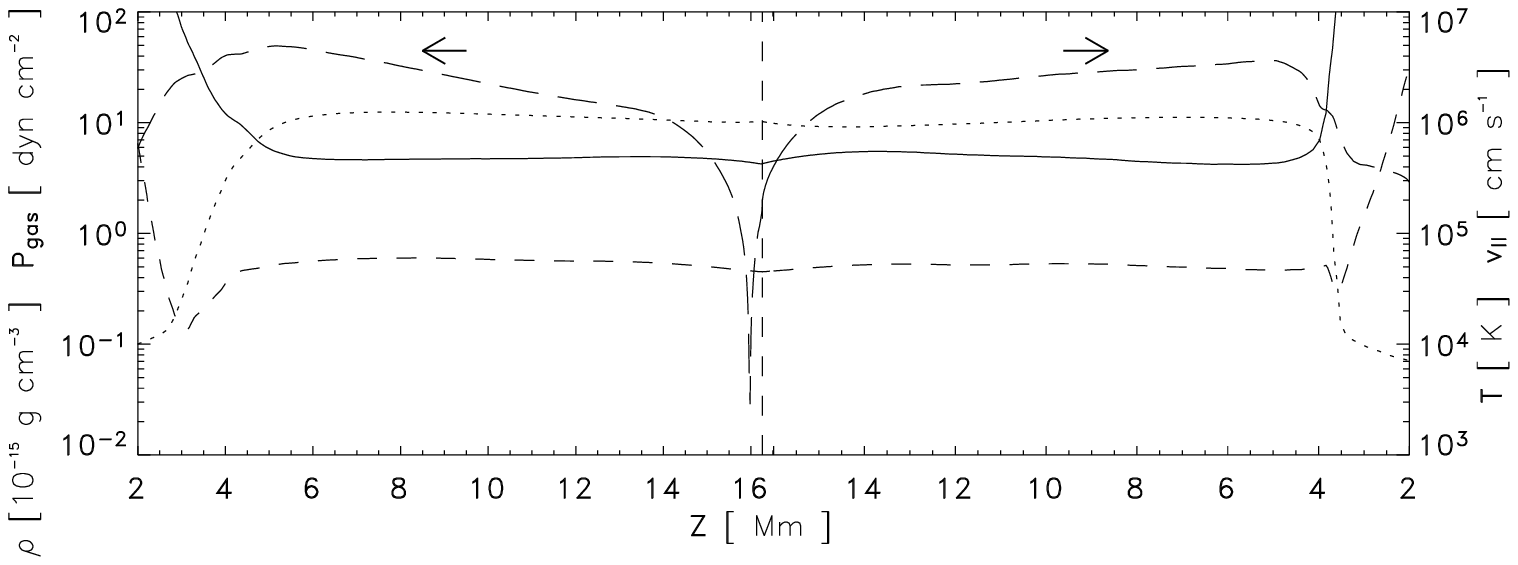}
           \label{fig:loop1-171-gas}
           \caption{\CAPVIII}
         \end{figure}
       }

       \newcommand{\FIGIX}{
         \begin{figure}[t]
           \figurenum{7}
           \plotone{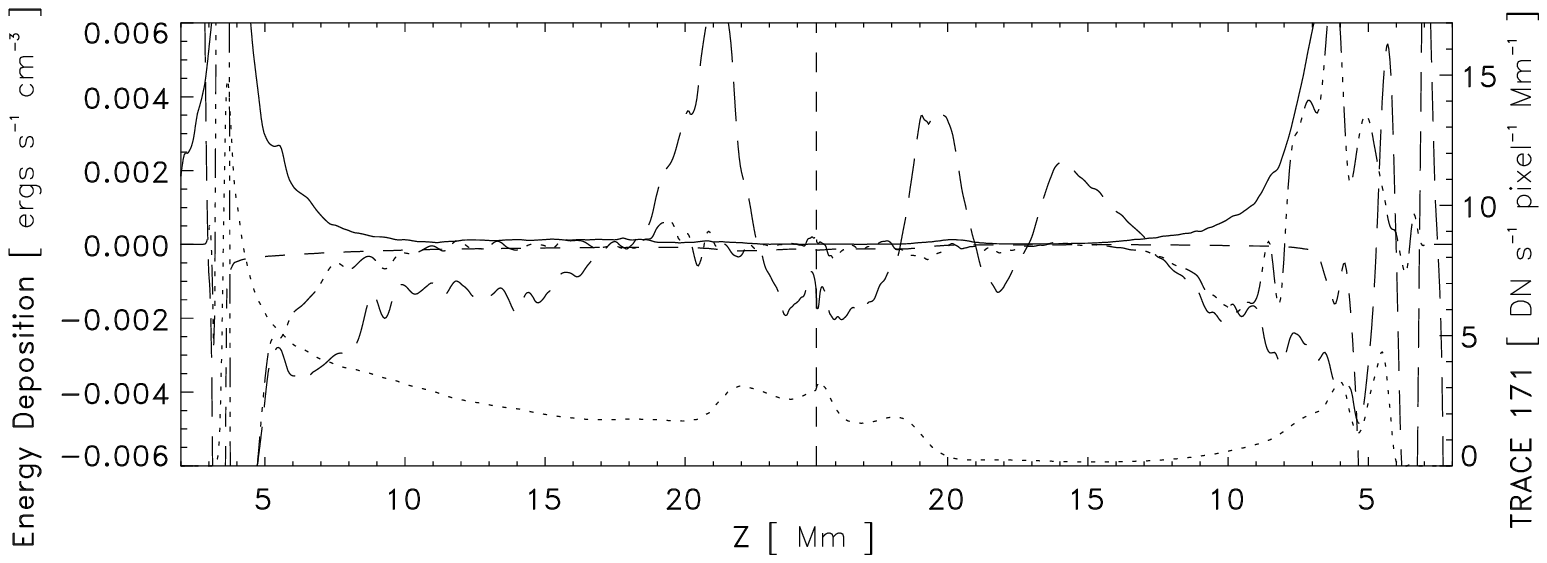}
           \label{fig:loop2-171-energy}
           \caption{\CAPIX}
         \end{figure}
       }
       
       \newcommand{\FIGX}{
         \begin{figure}[t]
           \figurenum{8}
           \plotone{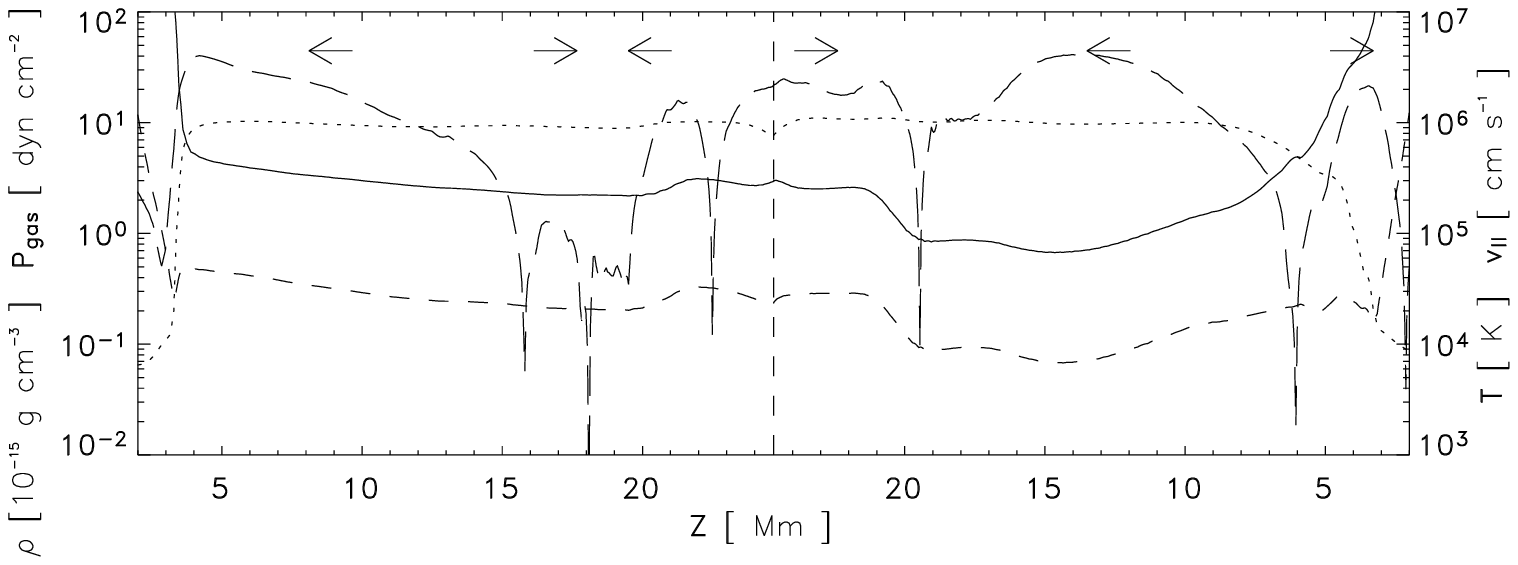}
           \label{fig:loop2-171-gas}
           \caption{\CAPX}
         \end{figure}
       }
       
       \newcommand{\FIGXI}{
         \begin{figure}[t]
           \figurenum{9}
           \plotone{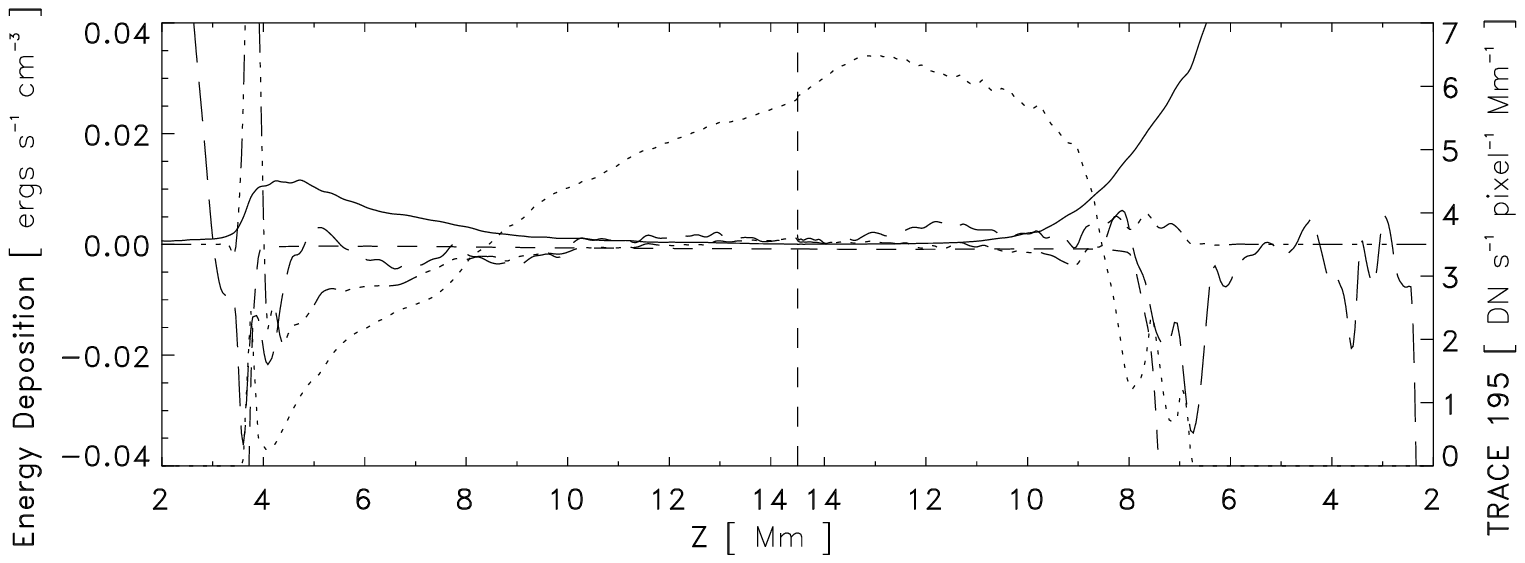}
           \label{fig:loop3-195-energy}
           \caption{\CAPXI}
         \end{figure}
       }
       
       \newcommand{\FIGXII}{
         \begin{figure}[t]
           \figurenum{10}
           \plotone{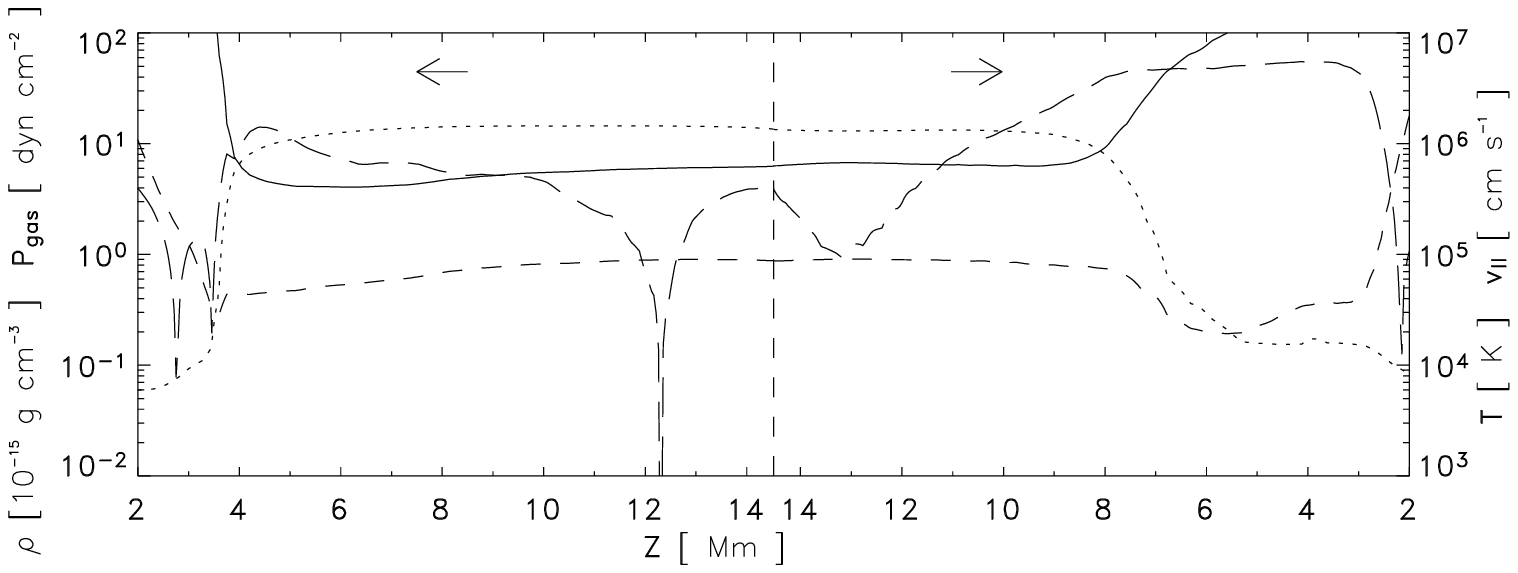}
           \label{fig:loop3-195-gas}
           \caption{\CAPXII}
         \end{figure}
       }

       \newcommand{\FIGXIII}{
         \begin{figure}[t]
           \figurenum{11}
           \plotone{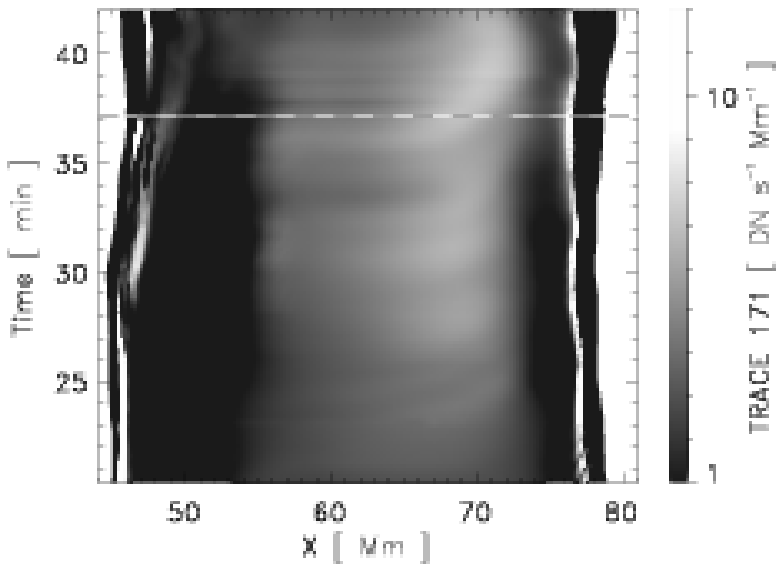}
           \label{fig:time-evo-tr171}
           \caption{\CAPXIII}
         \end{figure}
       }

       \newcommand{\FIGXIV}{
         \begin{figure}[t]
           \figurenum{12}
           \plotone{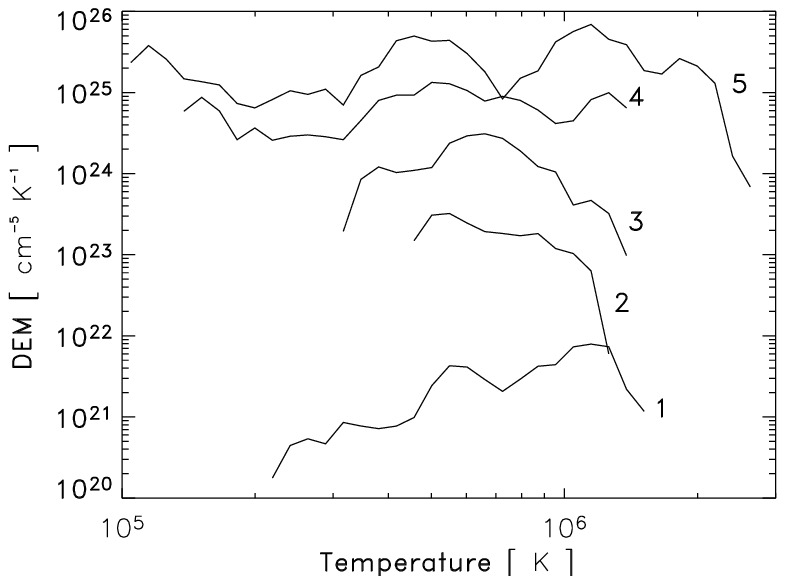}
           \label{fig:DEMtr171}
           \caption{\CAPXIV}
         \end{figure}
       }

       \newcommand{\FIGXV}{
         \begin{figure}[t]
           \figurenum{13}
           \plotone{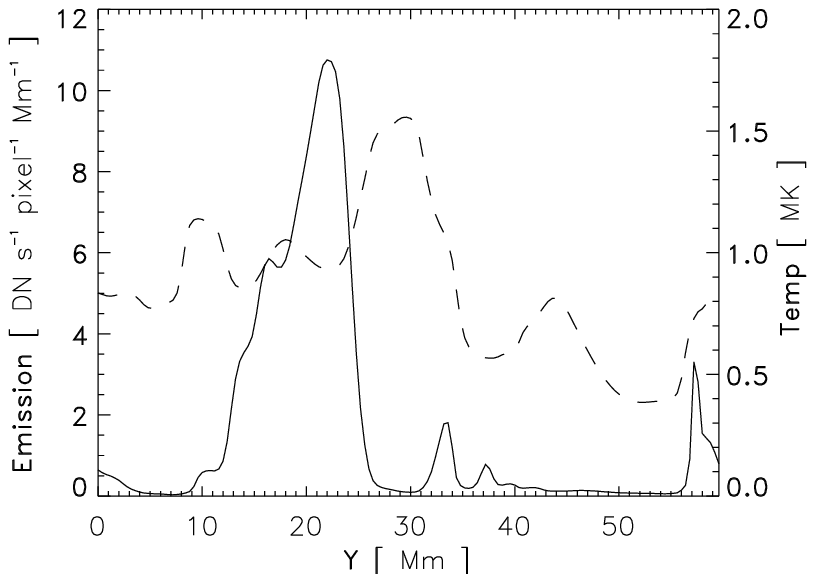}
           \label{fig:loop_point3}
           \caption{\CAPXV}
         \end{figure}
       }

\else
 
        \newcommand{\FIGI}{
          \begin{figure}[t]
            \figurenum{1}
            \centerline{\includegraphics[width=8.2cm]{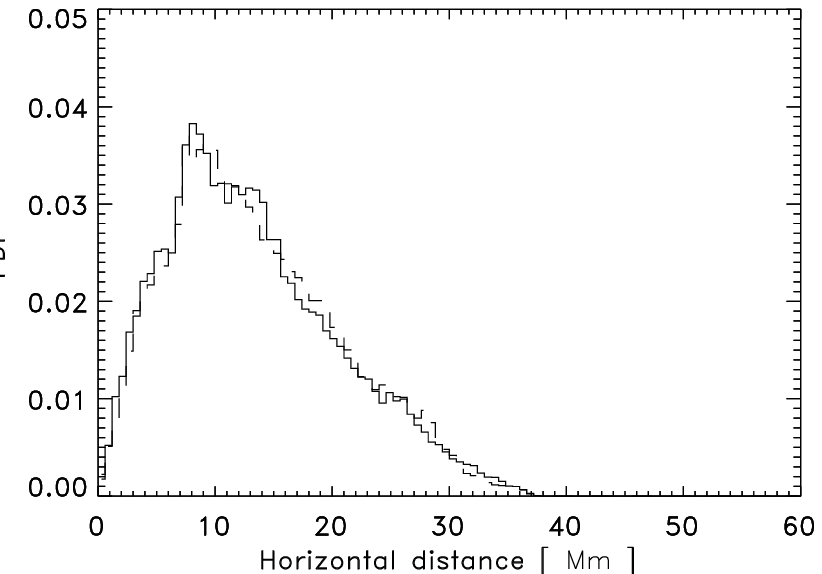}}
            \caption{\CAPI}
            \label{fig:connection1}
          \end{figure}
        }

        \newcommand{\FIGII}{
          \begin{figure}[t]
            \figurenum{2}
            \centerline{\includegraphics[width=8.2cm]{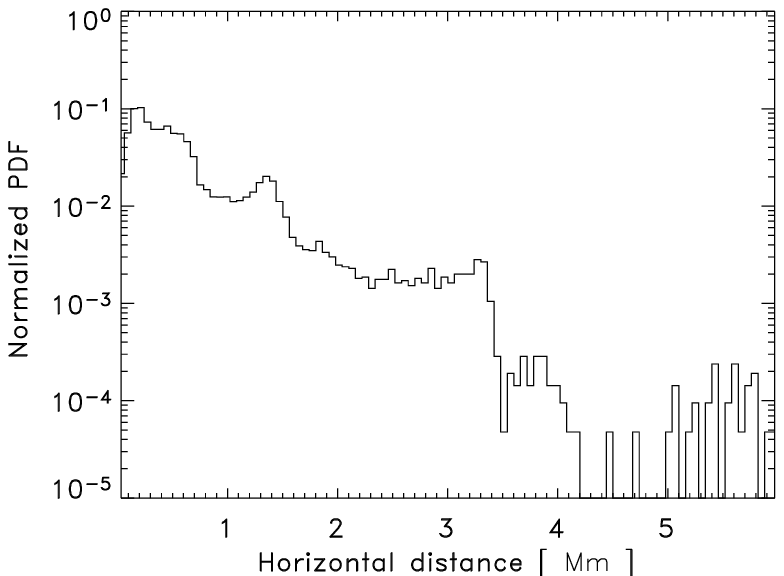}}
            \caption{\CAPII}
            \label{fig:connection2}
          \end{figure}
        }

        \newcommand{\FIGIII}{
          \begin{figure}
            \figurenum{3}
            \centerline{\includegraphics[width=8.2cm]{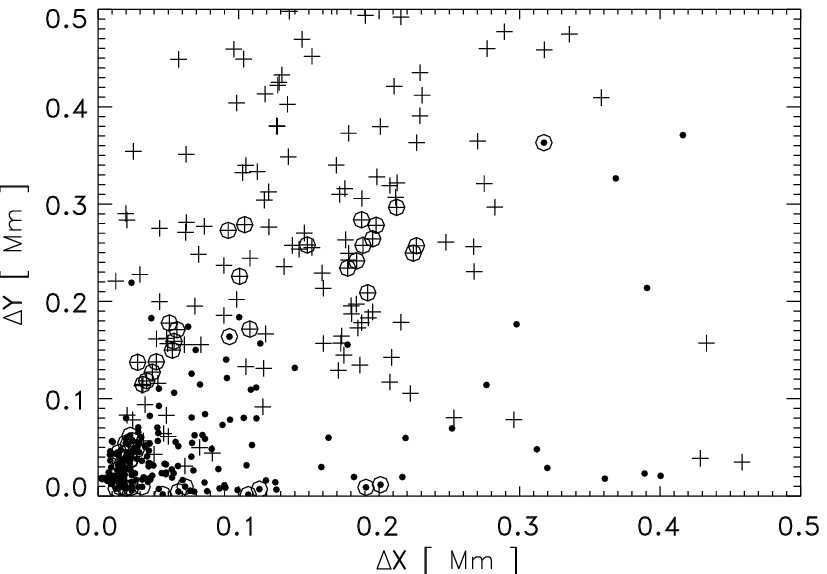}} 
            \caption{\CAPIII}
            \label{fig:footpoints}
          \end{figure}
        }

        \newcommand{\FIGVI}{
          \begin{figure}[t]
            \figurenum{4}
            \centerline{\includegraphics[width=8.2cm]{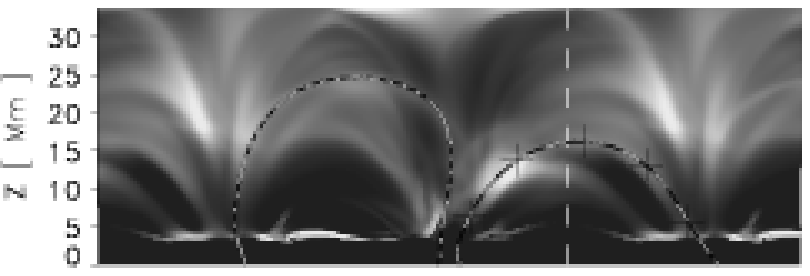}} 
            \centerline{\includegraphics[width=8.2cm]{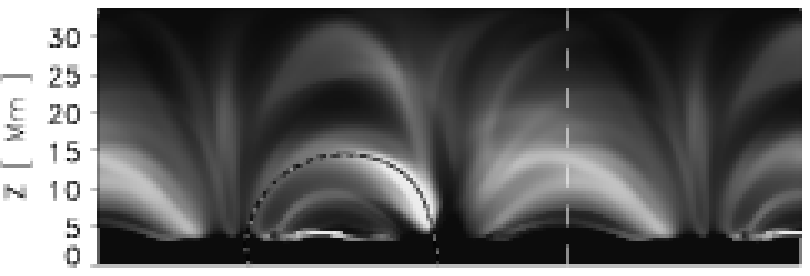}}
            \centerline{\includegraphics[width=8.2cm]{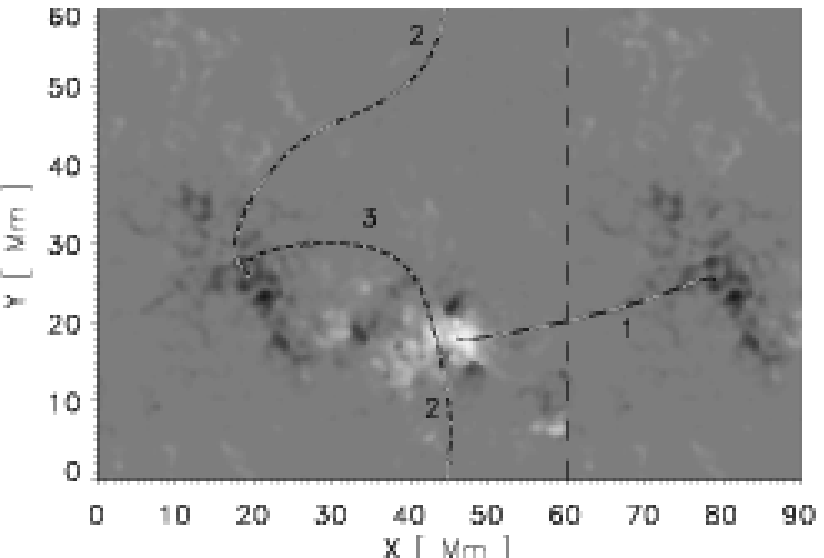}}           
            \label{fig:magneto70}
            \caption{\CAPVI}
          \end{figure}
        }

        \newcommand{\FIGVII}{
          \begin{figure*}[t]
                \figurenum{5}
                \centerline{\includegraphics[]{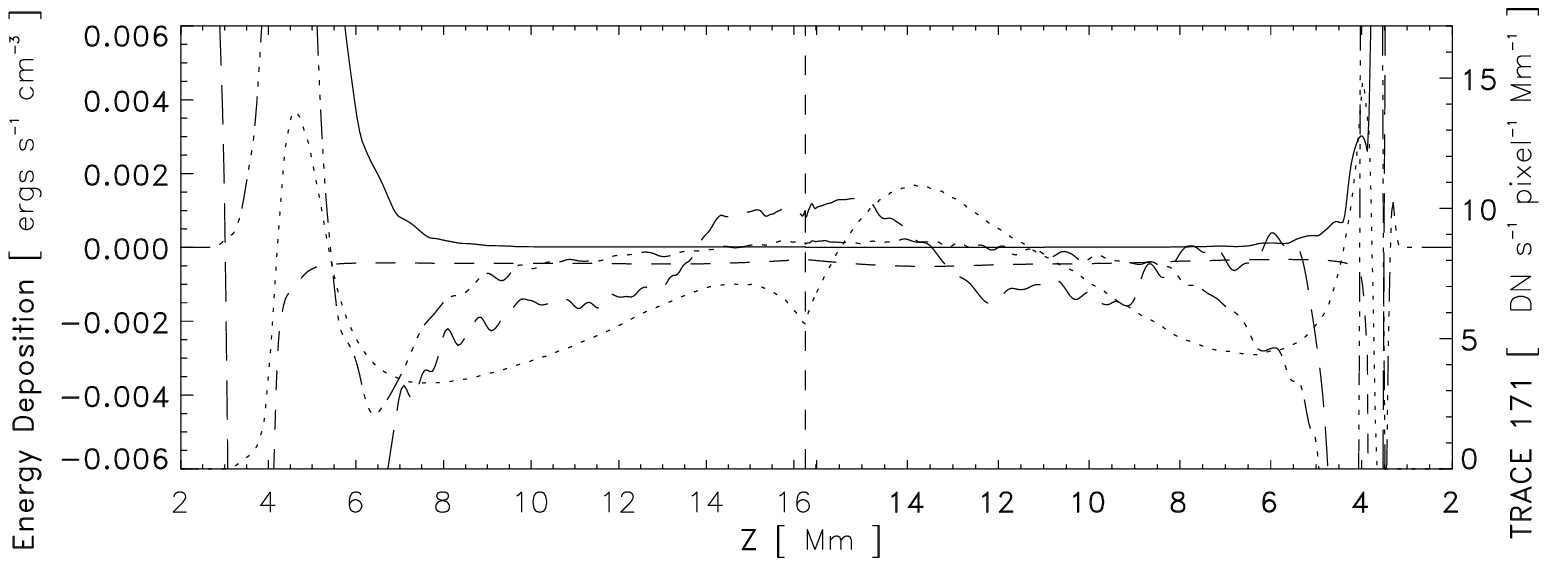}}
                \caption{\CAPVII}
                \label{fig:loop1-171-energy}
                \end{figure*}
        }

        \newcommand{\FIGVIII}{
          \begin{figure*}[t]
            \figurenum{6}
            \centerline{\includegraphics[]{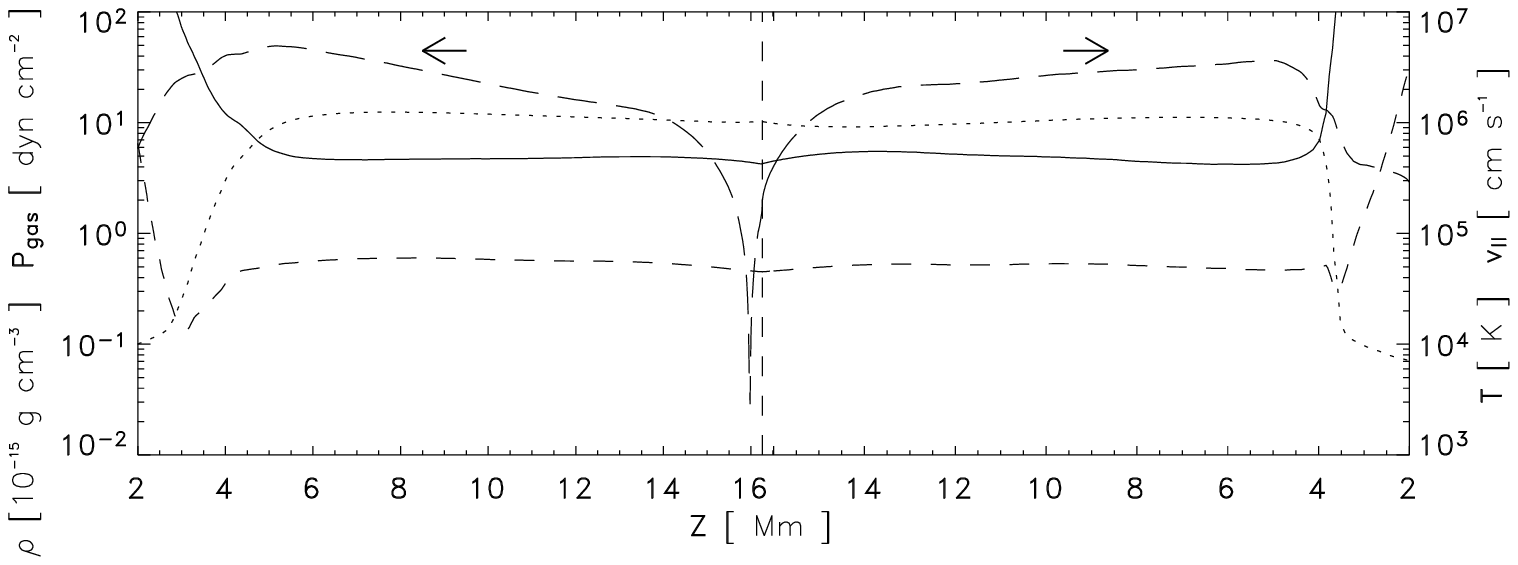}}
            \caption{\CAPVIII}
            \label{fig:loop1-171-gas}
          \end{figure*}
        }

        \newcommand{\FIGIX}{
          \begin{figure*}[t]
            \figurenum{7}
            \centerline{\includegraphics{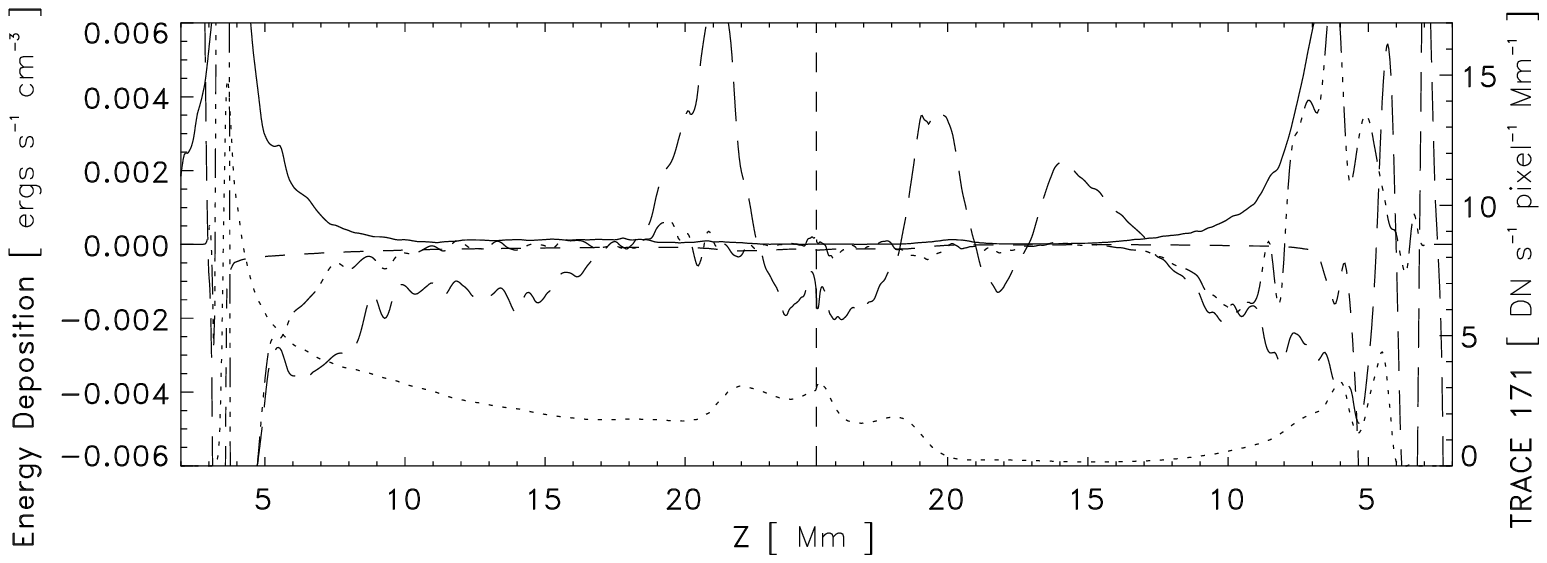}}
            \caption{\CAPIX}
            \label{fig:loop2-171-energy}
          \end{figure*}
        }

        \newcommand{\FIGX}{
          \begin{figure*}[t]
            \figurenum{8}
            \centerline{\includegraphics[]{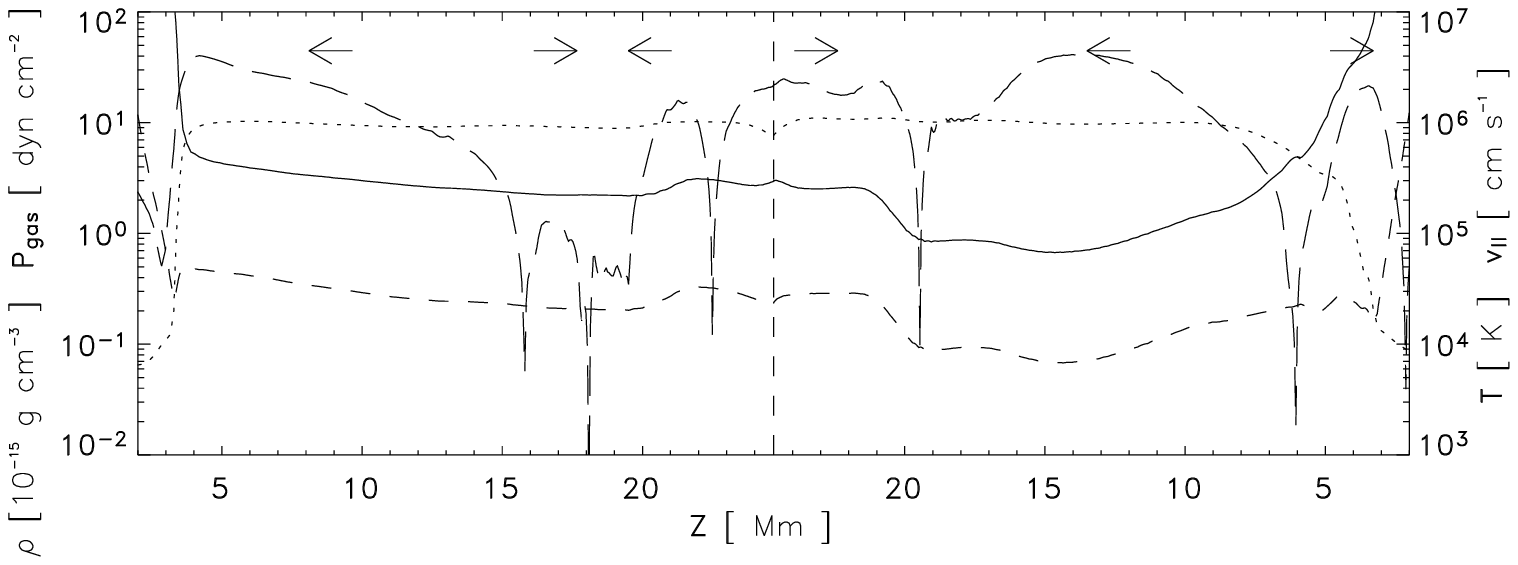}}
            \caption{\CAPX}
            \label{fig:loop2-171-gas}
            \end{figure*}
          }

        \newcommand{\FIGXI}{
          \begin{figure*}[t]
            \figurenum{9}
            \centerline{\includegraphics[]{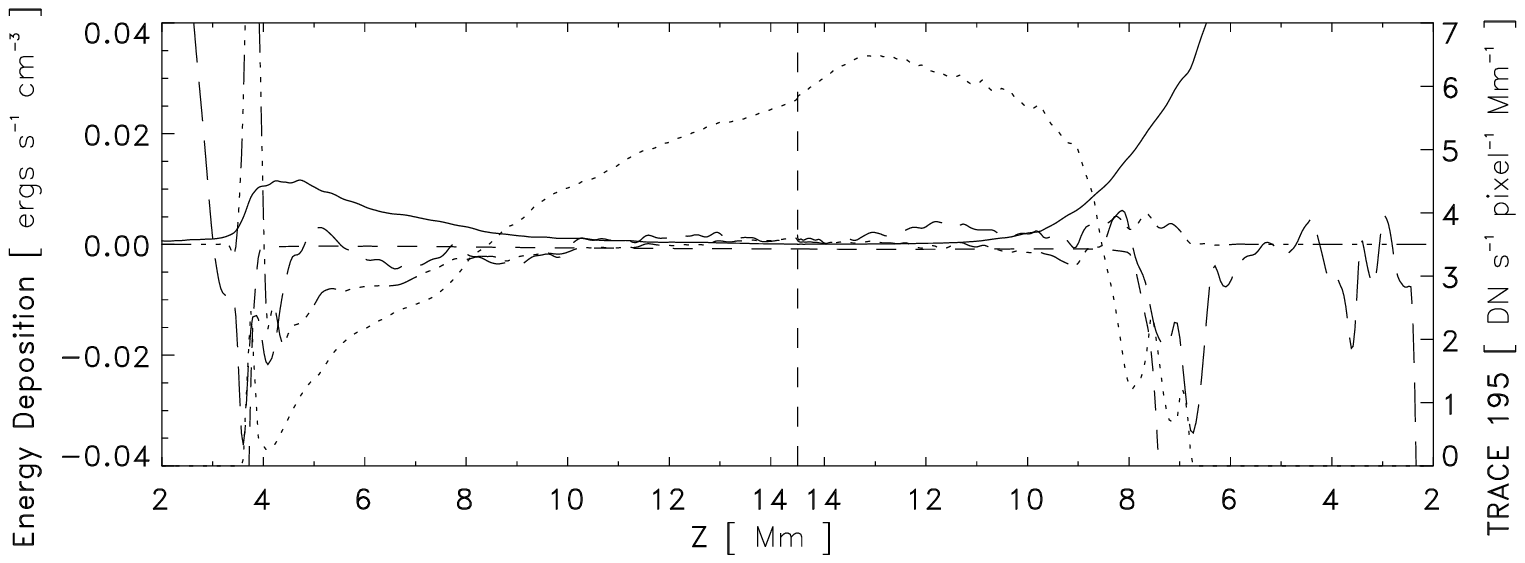}}
            \caption{\CAPXI}
            \label{fig:loop3-195-energy}
          \end{figure*}
        }

        \newcommand{\FIGXII}{
          \begin{figure*}[t]
            \figurenum{10}
            \centerline{\includegraphics[]{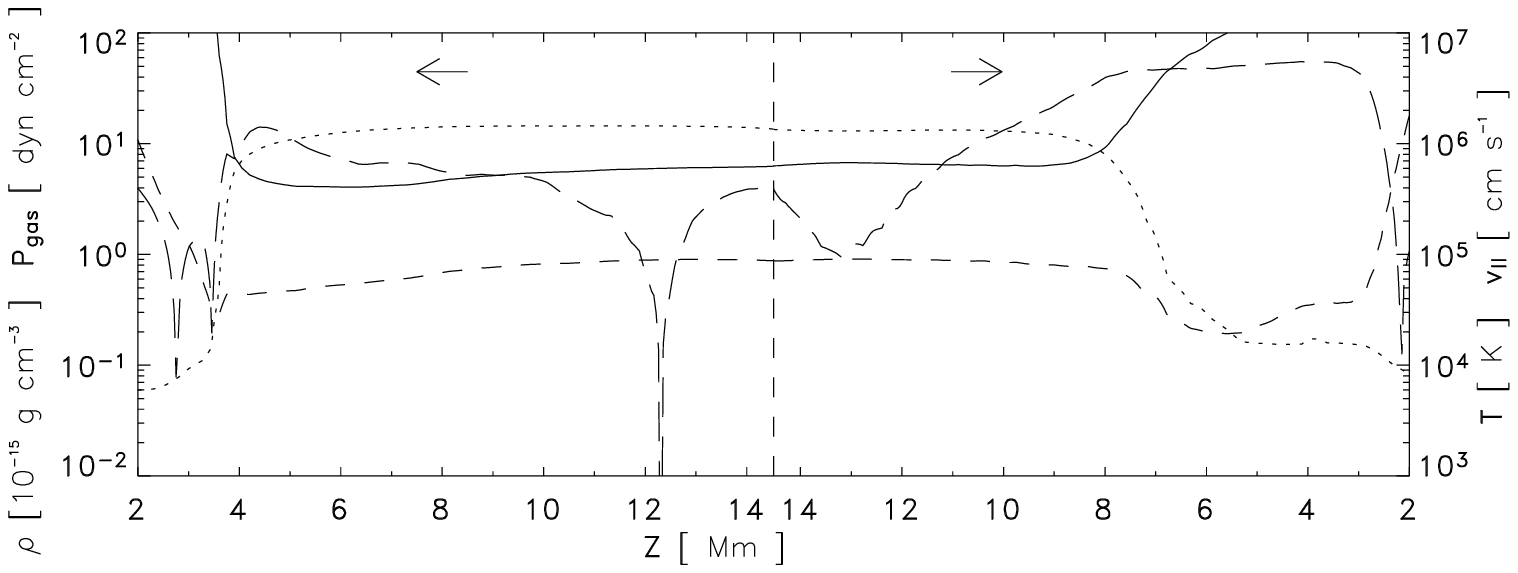}}
            \caption{\CAPXII}
            \label{fig:loop3-195-gas}
          \end{figure*}
        }

        \newcommand{\FIGXIII}{
          \begin{figure}[t]
            \figurenum{11}
            \centerline{\includegraphics[]{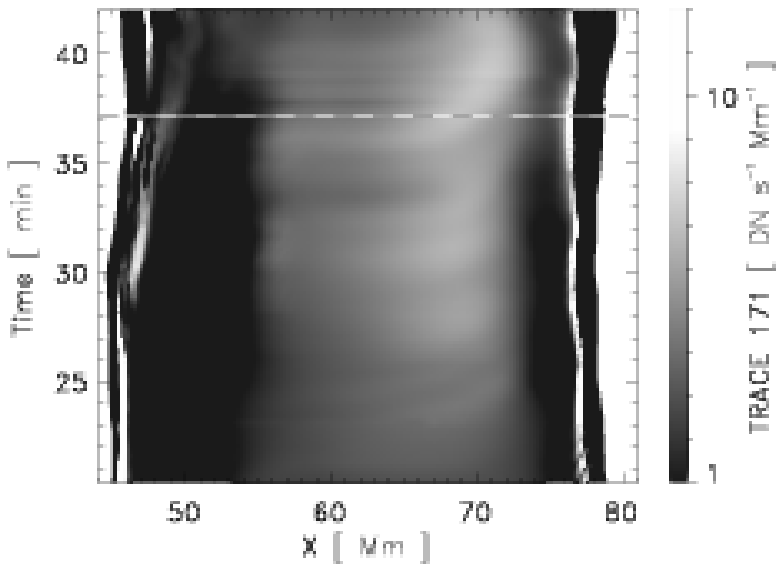}}
            \caption{\CAPXIII}
            \label{fig:time-evo-tr171}
          \end{figure}
        }

        \newcommand{\FIGXIV}{
          \begin{figure}[t]
            \figurenum{12}
            \centerline{\includegraphics[]{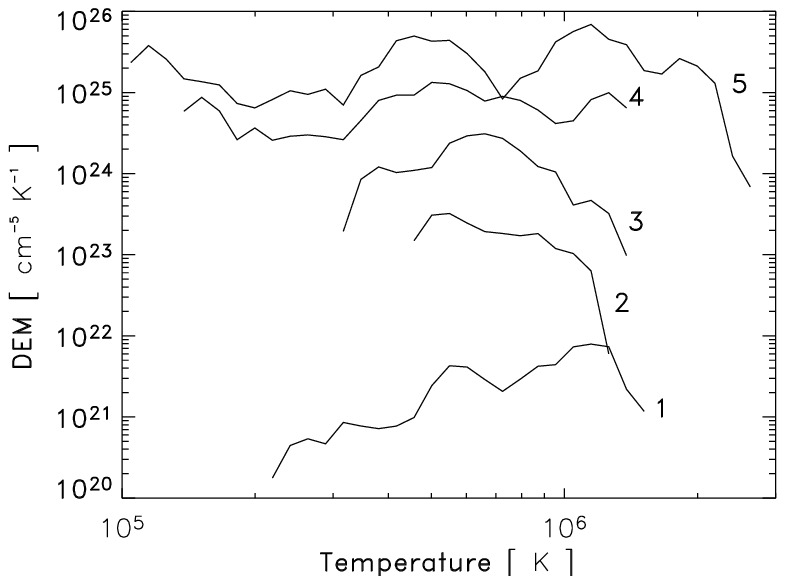}}
            \caption{\CAPXIV}
            \label{fig:DEMtr171}
          \end{figure}
        }

        \newcommand{\FIGXV}{
          \begin{figure}[t]
            \figurenum{13}
            \centerline{\includegraphics[]{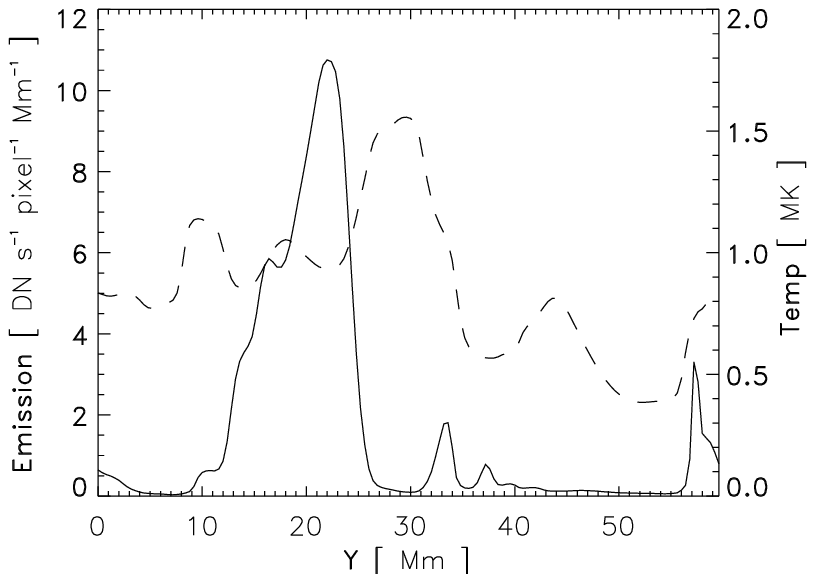}}
            \caption{\CAPXV}
            \label{fig:loop_point3}
          \end{figure}
        }

\fi

\newcommand{\Fig}[1]{Fig.\ \ref{#1}}
\newcommand{\Figure}[1]{Figure \ref{#1}}
\newcommand{\Figures}[1]{Figures \ref{#1}}
\newcommand{\Sec}[1]{Sec.\ \ref{#1}}

\begin{document}

\title{An Ab Initio approach to Solar Coronal Loops}

\ifnum\astroph=1
        \author{B.V. Gudiksen and {\AA}. Nordlund}
\else
        \author{Boris Vilhelm Gudiksen}
        \affil{The Institute for Solar Physics of the Royal Swedish
        Academy of Sciences}
        \affil{Albanova University Center, Stockholm Observatory\\ 10691 Stockholm\\ Sweden}
        \email{boris@astro.su.se}
        \and
        \author{{\AA}ke Nordlund}
        \affil{Astronomical Observatory}
        \affil{NBIfAFG, Copenhagen University}
        \affil{{\O}ster Voldgade 3\\1350 Copenhagen K\\Denmark}
        \email{aake@astro.ku.dk}
\fi

\begin{abstract}
Data from recent numerical simulations of the solar corona and transition
region are analysed and the magnetic
field connection between the low corona and the photosphere is found
to be close to that of a potential field. The fieldline to fieldline
displacements follow a power law distribution with typical displacements
of just a few Mm. Three loops visible in emulated Transition Region
And Coronal Explorer (TRACE) filters are
analysed in detail and found to have significantly different heating
rates and distributions thereof, one of them showing a small scale heating
event. The dynamical structure is complicated even though all the
loops are visible in a single filter along most of their lengths. None of
the loops are static, but are in the process of evolving into loops with
very different characteristics. Differential Emission Measure (DEM) curves
along one of the loops illustrate that DEM curves have to be treated
carefully if physical characteristics are to be extracted.
\end{abstract}

\section{Introduction}
Since the launch of the SOlar and Heliospheric Observatory (SOHO) and
later the Transition Region And Coronal Explorer (TRACE), observations
of EUV loops in the solar corona have become sufficiently good to provide
a testbed for coronal loop models. Earlier, \citet{Rosner+etal78} produced
scaling relations for loops with constant pressure, uniform heating
and constant cross section along their lengths, the so called RTV
scaling laws. These were later 
generalized by \citet{Serio+etal81}, including non-uniform heating,
and non-uniform pressure in the form of two scale
heights. It has though
become increasingly clear that even these hydrostatic models have problems
reproducing the majority of the loops observed in the EUV. 
\citet{Aschwanden+etal01} compared 41 loops observed with TRACE with
standard loop
models, and found that none of the observed loops could be fitted 
with the RTV scaling laws and only roughly 30 \% of the loops could be
explained by hydrostatic solutions with foot point heating. 

Recently
\citet{Winebarger+etal03} compared both EUV and X-ray loops and found
that long EUV loops are overdense by up to 3.4 orders of magnitude
compared to hydrostatic uniformly heated models. Furthermore, only
28 \% of the 67 loops were explainable by hydrostatic non-uniformly heated
models. This included X-ray loops with an assumed filling factor of unity,
which could in principle be reconciled with the models by
assuming a smaller filling factor, but this in itself would introduce other
problems. Reducing and interpreting EUV coronal loop observations are
complicated and cannot provide all the necessary
information. Obtaining reliable estimates of gas parameters 
and velocities for a whole loop 
demands spectral information only available by rastering, which
compromises the understanding of the dynamic nature of loops. 

Further problems
concerning proper background subtraction for both spectra and imaging
have been shown to have effects on the deductions made
\citep{Martens+etal02,Schmelz+etal03,DelZanna+Mason03}. Modelling of
observed loops consequently incorporates assumptions not necessarily
confirmed by observations. Common for most models is the 
assumption of a uniform loop cross section. This assumption is unchallenged
by observations since the measured expansion of the loops is severely
limited by the spatial resolution of existing instruments. Lately,
claims have been made that loops now appear to be resolved by TRACE
\citep{Testa+etal02}. At the same time, loops of cross section on the
order of a few pixels show signs of consisting of smaller structures
since estimates of cooling time, based on the evolution of the
intensity in two TRACE pass bands, do not agree with hydrostatic loop
models \citep{Warren+etal03}. 
Results from a 3D MHD model of the corona gives us the opportunity to
compare individual loops identified in the model with observations and
investigate if the often used approximations in loop models are valid.

The problems of modelling coronal loops will here be treated with 
data from the numerical simulations of
\citet{Gudiksen+Nordlund03a}. We will discuss the magnetic field state
and the validity of a potential field extrapolation of a magnetogram,
the assumption of constant circular cross sections for loops, as well
as looking at flows, energy balance and gas parameters for a few
single loops. The time evolution of one of these loops and the
DEM curves for a few points along the same
loop will also be discussed. Emission, Doppler shifts, and non-thermal
widths of a number of selected spectral lines are treated in a
separate paper \citep{Peter+etal04}.

\section{Model Description}
The numerical simulation of the solar corona is described in detail
in a separate paper \citep{Gudiksen+Nordlund03a} and is only
introduced briefly here.
The model is based on an MHD code that incorporates relevant physics,
including a radiative cooling function and Spitzer 
conductivity. The model spans $60\times60$ Mm$^2$ of the
solar surface, and stretches 37 Mm up into the
corona from the photosphere. A model for the solar photospheric
velocity field stresses the 
magnetic field which initially is a potential extrapolation of an
MDI/SOHO high resolution magnetogram of AR 9114, scaled to fit inside
the box. The initial thermal stratification is a FAL-C \citep{Fontenla+etal93} 
atmosphere in the
photosphere and chromosphere with an isothermal 1 MK corona above the
transition region. The temperature in the lower, optically thick
atmosphere is forced toward the initial temperature profile on a
typical timescale of \mbox{0.1 s} in the photosphere and decreasing as
$\rho^{2/3}$, making it unimportant in the transition region and
corona. To check the effect of the chromospheric 
stratification on the corona a ``cool'' chromosphere
with no average temperature increase \citep[see][and references
therein]{Carlsson+Stein02} was also used. 
The choice of chromospheric model is not of great importance for the
loop structures, and only the model with the standard FAL-C
chromosphere will be discussed here.

\ifnum\astroph=1
\FIGI
\fi
     
\section{Results}
The definition of a loop is often unclear, and can lead to a number of
misconceptions, so we will explain what we mean by the term ``loop''
and what the consequences are. A loop observed in a narrow wavelength
band is {\em{defined}} by the plasma emitting in that certain wavelength
band. In principle that definition is independent of the magnetic field, but
in the 
solar corona it can, to a very good approximation, be assumed that the near
isothermal plasma seen emitting is caught in a flux bundle, because of the
efficient thermal conduction along the
magnetic field. Often magnetic field lines are used to trace the
magnetic field, but it is important to emphasize that a field line
is not a physical entity, it is simply a line following the direction
of the magnetic field. It is therefore not affected by changes in the
local magnetic field strength, and cannot be followed in time.   
The magnetic field in the solar 
corona is space filling and it is rarely, if ever, possible to identify a loop
from the magnetic field alone, without the information from narrow
band imaging instruments. Figures where loops are shown as a number of
field lines are made by a subjective choice of field lines that along
their lengths have gas emitting in a particular wave length band.  
Consequently, loops are the result of plasma being near isothermal, 
and not the effect of the magnetic field being special in the volume defined
by the emitting plasma. A visible loop is a result
of the heating history within the flux bundle in which the emitting
gas is caught and not a result of the magnetic field in the flux bundle 
being very much different from the magnetic field nearby. 

It is often assumed
that the emitting gas and the same magnetic flux are involved in a long
lived loop, and even though this is often the case, it need not be. In
a 3D simulation such as this, one is fortunate enough to able to choose
between identifying loops by following the magnetic field, and
identifying them from a high level of emission. In case of time
evolution following a loop based on the magnetic field is not
possible---at least not in a unique way---so instead one has to 
follow a plasma parcel through time.  Only to the extent that the
field is ``frozen in'' does this mean that the same flux is passing 
through the plasma parcel.

\ifnum\astroph=1
\FIGII
\fi

\subsection{Loop connection from photosphere to corona}\label{sec:res1}
Observed polarisation signals from the Sun usually
originate from a relatively thin layer in the photosphere \citep[but
see recent results from ][]{Solanki+etal03}. The 
connection to a 3D magnetic field is then usually made by using the observed
photospheric magnetogram and extrapolating by a method producing
either a linear force free or 
potential field. With the movements of the magnetic foot points in the
solar photosphere that method must be questioned. To investigate 
whether this method provides a good approximation we have compared
the horizontal 
distance a fieldline traverses from the photosphere to a predetermined height
just above the transition region. The magnetic fieldlines are followed
from a horizontal layer just above the transition region where they are
distributed evenly over the whole layer. \Figure{fig:connection1} is a
histogram showing the distribution of the distances for a typical snapshot and
for a potential extrapolation of the vertical magnetic field in the
photosphere. Even though the difference between these distributions is
very small, the distance between the foot point of a
field line passing through a point in the low corona obtained from the
simulation and the assumption of a potential magnetic field may be
very large for individual fieldlines. \Figure{fig:connection2} shows
the distance between 
foot points of a fieldline passing through the same point in the lower
corona from the potential extrapolation and from the snapshot. Around
Quasi--Separatrix--Layers (QSLs) \citep{Priest+Demoulin95} the
computed distances are generally very large,
because the shuffling of the fieldline foot points make the
QSLs move around in the atmosphere, making fieldlines starting at
the same geometrical point end up at large distances from each
other. Fieldlines through such points are not
included in \Fig{fig:connection2}. The PDF is consistent with a 
power law but since we do not have the full range of driving scales
included, and because the simulation has run for only one turnover time of
the largest driving scale it is plausible that we would get a high
distance tail if the simulation had run for a longer period. For the
majority of the fieldlines followed here the difference in distance is modest,
on the order of a few Mm. This is a small change in connectivity, and
in spite of the maximum driving 
scale being smaller than the super granular scale typically used,
substantial heating is produced. 

\ifnum\astroph=1
\FIGIII
\fi

Fieldlines that do not reach above the transition
region are disturbed the most. This is partly because the higher densities 
in the chromosphere give low propagation speeds for disturbances,
and thus allow larger distortions.
Correspondingly, roughly 90 \% of the dissipated energy is injected
in the lower atmosphere and not in the corona \citep{Gudiksen+Nordlund03a}. 
Large scale shear in the corona can only be
created by a large scale persistent photospheric velocity field,
if it is not already a property of the
magnetic field at the time of emergence.

\subsection{Loop cross section}
Loops have generally been assumed to be of constant cross section at
all heights in the corona, based on the observations from SOHO
and TRACE, where there seems to be no increase in cross section with
height, contrary to what would be expected for a potential field.
Implicitly, to at all reach conclusions about cross sections from such
observations, it has been assumed in previous works that cross
sections of flux surfaces are roughly circular. For the
heating mechanism evident in this work, such an assumption is 
highly unlikely. To achieve heating, magnetic fieldlines have
to be at an angle relative to each other at the heating
location. Identifying a loop by its  EUV or X-ray emission and therefore
indirectly by its heating history, one can be certain that the field
lines in such a loop are not aligned. 
The cross section of a flux surface that is circular at one point
along a loop will not remain circular in a tangled fieldline geometry. 

\ifnum\astroph=1
\FIGVI
\fi

\ifnum\astroph=1
\FIGVII
\FIGVIII
\fi

\Figure{fig:footpoints} shows the horizontal width of loops that
have a circular-loop cross section at their tops of \mbox{0.8
Mm}. Loops with a wide variety of maximum heights have been chosen, but
there is no obvious dependence of the changes in cross section
on the maximum height. It
is clear that the loops have significantly smaller cross 
sections in the photosphere, and that they are non-circular. 
When moving further
up into the transition region, the loops become wider, some having
expanded by a factor of ten but generally the expansion is not identical in
all directions. That makes the traditional ``wine-glass'' picture of
the magnetic field in the transition region a bad approximation, as
\citet{Schrijver+Title03} have pointed out after modelling of network
flux concentrations. There is a tendency for bright TRACE 
loops to be very asymmetric in the photosphere. 
It is clear that if these loops are assumed to have circular cross
sections at their tops, few (if any) have circular cross sections at a
height corresponding to the transition region. Therefore neither cross 
section nor cross section area are conserved along any of the loops we have
followed. In general the foot prints of the loops shown in
\Fig{fig:footpoints} are wrinkled and not geometrically simple, and their
area cannot be approximated by product of $\Delta X$
and $\Delta Y$.  We expect the foot prints to increase in complexity with
increasing resolution.
 
It should
be pointed out that the loops followed here are not chosen on the
basis of a visible characteristic, but are simply based on choosing 
random points in the corona, whereafter a set of fieldlines around
each point are followed to the photosphere. This means that not all
the loops treated here are 
representative of the observed cross sections of TRACE loops, but
lend evidence to the fact that bright loops in general do not
have circular cross sections along their whole length. In order to
probe the true cross section of 
loops, the spatial resolution would have to be raised by at least a
factor of two, to include velocity gradients produced by flow scales
in the photosphere that the resolution here cannot resolve.

\ifnum\astroph=1
\FIGIX
\FIGX
\fi

\subsection{Loop Heating}
Loops have in general highly time dependent heating profiles, and the
heating is independent in the loop legs. It is thus only possible to 
define a general heating profile if a large number of loops are used. We obtain
an average heating profile that above the transition region
decreases exponentially, with a scale length of \mbox{$\sim6$ Mm}.
The height dependence of the heating is
highly variable from loop to loop. The scale height behaviour emerges since the
heating is proportional to the current squared and because the magnetic field
in the corona is close to force free, which makes the current along each 
fieldline proportional to the magnetic field.
In addition to being close to force free, with the property
$\vec{J}\propto \vec{B}$, the magnetic field also does not deviate much
from a potential field (cf.\ \Sec{sec:res1}).
The overall magnetic field strength thus decreases roughly exponentially
with height, with a scale height determined by the distance between the 
dominating magnetic polarities. 
In the present periodic model the shortest distance between the two main 
polarities is roughly half the box size, so at coronal heights the
wave number corresponding to the box size $k_0$ dominates.
Assuming that the magnetic field drops off with height as a potential
field we thus obtain
\begin{eqnarray}
Q(z)&\propto& J^2(z)\propto B^2(z)\propto \left[\exp{\left(-k_0\,
 z\right)}\right]^2\nonumber \\
&\propto& e^{-2k_0\,z} ,
\end{eqnarray}
and so the estimated heating scale height is $s_H=1/2k_0\sim5.0$
Mm, consistent with what we actually measure.  

In terms of the half loop length $L$ of assumed semicircular loops 
\citep[cf.\ ][]{Aschwanden+etal01}, our estimated heating scale height is
\begin{equation}\label{eq:llfrac}
{s_H \over L} = {1 \over 2 k_0 L} = {2 \over \pi^2} \approx 0.2 ~, 
\end{equation}
where the wave number $k_0$ and the half loop length $L$ have been related 
to the distance between polarities $d$ through $k_0 = \pi/d$ and 
$L = \pi d/4$.  This is consistent with \citet{Aschwanden+etal01}, 
who find $s_H/L = 0.2 \pm 0.1$.

\subsection{Loop stratification and dynamics}
Loops encountered in this model are generally not in equilibrium and
a significant fraction of the visible loops identified in the TRACE 171
filter are not aligned with the magnetic field. A possible scenario for
such a loop to develop is when one of two connected polarities has a
velocity gradient in the direction connecting the two polarities. This
implies that one polarity is drawn out along a line perpendicular to
the line initially connecting the two polarities, which will create a
shear in the field connecting the two. In general the maximum of the
shear will not follow a particular fieldline connecting the
two polarities. If the shear happens on a short timescale the maximum shear
locations will enter the narrow TRACE filters at the same time,
showing a bright region that is not connected by a single loop
bundle. 

\ifnum\astroph=1
\FIGXI
\FIGXII
\fi

Three loops are traced in \Fig{fig:magneto70}, where loop 1 and
loop 2 are selected for their brightness in the emulated TRACE 171
filter (\Fig{fig:magneto70}, top) while loop 3 is 
selected for its brightness in the emulated TRACE 195 filter
(\Fig{fig:magneto70}, middle).


Loop 1 can be identified through its whole length in the emulated
TRACE 171 filter. It is an example of a steady relatively cold
loop and seems to be in 
a cooling phase. It shows low steady velocities down both of its legs of at
most a few tens of $\textrm{km s}^{-1}$. The energy
balance and the gas parameters of the loop are shown in 
\Figures{fig:loop1-171-energy} and \ref{fig:loop1-171-gas},
respectively. The heating is growing smoothly towards the photosphere,
except for a small bump on the right leg, which could be an impulsive
heating event. The convective energy term advects energy from the
coronal part of the loop to the denser lower atmosphere, were it is
deposited. The Spitzer conductivity shows the characteristic
pattern of moving energy from the hot corona and pumping it down into
the chromosphere. The radiative cooling is insignificant and shown
here without quenching, making it rise sharply when the density
increases. This loop is very quiescent except for the bottom of the
right leg, which shows large fluctuations in the convective term, due
mainly to the large densities, and not so much large velocities. The
temperature, density and pressure are very close to constant in most
of the corona. There is a slight decrease in temperature, which is
large enough to have an effect on the emission in TRACE 171 filter. 

Loop 2 is shown in \Figures{fig:loop2-171-energy} and
\ref{fig:loop2-171-gas}. This loop is an example of a dynamic loop
that, in spite of the intermittent velocities present in the loop, is
still visible throughout most of its length. This loop has more
uneven heating than loop 1. The velocities present in this loop are the
effect of the heating and velocities perpendicular to the loop. This
loop is much longer than loop 1, even though this is not obvious 
from \Fig{fig:magneto70} (middle), the top view shows that this loop
is much longer than loop 1. Projection effects such as this one
are quite common, and lead to confusing and at times misleading
conclusions about a particular loop. The special geometry of the loop,
with a flat loop top, makes the gradients in velocity in
\Fig{fig:loop2-171-gas} seem strong because the velocity is plotted
against height. In spite of the gradients being smaller than
what they seem from \Fig{fig:loop2-171-energy}, there is still a
significant energy reorganisation in this loop. The convective energy
term is dominating everything else in the corona produced by
velocities of up to 30 km s$^{-1}$ in the right leg. The left leg has
a distance over which the velocities are on the order of \mbox{1
  km s$^{-1}$}. 
The temperature along the loop is close to 
constant through the whole corona, while the pressure and density are
larger in the left leg, with a noticeable increase in density at the
loop top. It is because the loop top is flat that it is possible for
this amount of mass to have accumulated at 
the loop top. It could have happened as the result of powerful heating
at both legs, which has now turned off. The evaporated gas would then
accumulate at the loop top, and now the loop is cooling down. The low
pressure of the right leg is at this point in time not 
able to support the dense gas at the loop top, which is now falling
down the right leg. At the same time there is still material coming up
the right leg, and a large amount of convective energy is deposited
where the two flows meet. The left leg also has material coming down from the
top, but this is slowed down high in the corona, with a corresponding
effect on the convective energy term. Below this point gas seems to be
free falling due to the very low magnetic heating in the left leg. 

Loop 3 is somewhat hotter than loop 1 and 2, and shows up
most prominently in TRACE 195. \Figures{fig:loop3-195-energy} and
\ref{fig:loop3-195-gas} show
that the heating in the right leg is more than a factor of ten higher
than in loop 1 and 2, and furthermore is much larger in the right leg,
compared to the left. From \Fig{fig:magneto70} one can see that the
right leg is rooted in the main polarity, with a magnetic field
strength roughly ten times higher than the left leg. Because of the
high pressure in the right leg, the transition region is also at
different heights in the two legs. In the left leg, the transition
region is at a height of 3.5 Mm while in the right leg it is at 7
Mm. As already mentioned the heating is in general dependent on the
magnetic field strength and the local height of the transition
region. Because of the large differences in magnetic field strength
and local height of the transition region, the heating is much
stronger in the right leg than in left. 
In spite of the 
powerful heating in the right leg, the radiative cooling close to
balances the powerful heating. The pressure in the right leg is
building, and is now higher than at the top of the loop. Nevertheless,
the momentum of the falling gas is still large enough that the increasing
pressure gradient is not yet able to break the fall.

\section{Time evolution}
Loops observed with TRACE and SOHO/EIT can be long lived. Some show
life-times much longer than dynamical, cooling and conductive time
scales. In following a loop in time we have attempted to follow an
emitting plasma-parcel which initially is caught in a flux
bundle. The plasma parcel is then followed to the next snap shot, and
here the magnetic field is used as a tracer for the rest of the
loop. In a 3D simulation like the one  
discussed here, following a plasma-parcel from one snapshot to the
next is theoretically simple. But following a plasma parcel over a
long time may require sub-resolution precision, because errors
in position accumulate from snapshot to snapshot. Unless the plasma
parameters change abruptly there is no way to estimate if an error 
has been made when following a plasma parcel over long times. In
spite of these problems we believe we have managed to follow the time
evolution of loop 1 with good precision for more than 20 minutes. The
emission in TRACE 171 is shown in \Fig{fig:time-evo-tr171}, in units 
of DN s$^{-1}$ pixel$^{-1}$ Mm$^{-1}$
(making the actual photon count rate depend on the thickness of the
loop as well as on the background and foreground). One resolution element,
emitting at the 
level of most of the upper part of the loop, makes up ~10 \% of the
typical level of emission from a loop seen in
\Fig{fig:magneto70}. The loops in this simulation are in general
thicker than in the solar corona, because of the limited resolution in the
photosphere, which sets the minimum driving scale. 
After the expansion of the magnetic field through the
chromosphere and transition region, this makes the loops in the
simulated corona thicker than on the Sun. In this simulation, loops
generally are on the order of 6-8 grid cells thick or \~3 Mm. In spite of
the very dynamic nature of the loop 1, it manages to stay bright along
most of its length during the whole 20 min time period. The
dark left leg of loop 1 is caused by a too high temperature, moving
that leg outside the temperature response function of the TRACE 171
filter. The reason that the leg is still apparent in
\Fig{fig:magneto70} (top) is the large amount of other loops
originating in the same area. \Fig{fig:time-evo-tr171} clearly shows 
that it is possible to have long lived, non static loops in a corona
produced by a highly time dependent DC heating mechanism. 

\section{Differential Emission Measure}
\ifnum\astroph=1
\FIGXIII
\fi
DEM can be constructed from
observations of a number of lines, essentially giving a measure for
the amount of emitting gas at a certain temperature. DEM can also be
constructed by comparing emission in a number of narrow filters, such 
as the ones used by EIT and TRACE. Constructing DEM
curves on the basis of narrow band observations is however an ill posed
problem, and whether any trustworthy information can be gained from
such exercises has been the subject of intense
discussion\citep{Schmelz+etal01,Martens+etal02,Aschwanden02}. Creating
DEM curves from a 3D simulation is more straight forward. The DEM can
be defined as 
\begin{equation}
\textrm{DEM}=n_e\frac{dh}{dT}\, , 
\end{equation}
which, when defining a volume emission measure ($\textrm{EM}_V$) as
\begin{equation}
\textrm{EM}_V(T)=\int_Vn_e^2dV \, ,
\end{equation}
 makes it possible to find the DEM from the simulation simply as 
\begin{equation}
\textrm{DEM}=\frac{\textrm{EM}_V}{A\,\Delta T}
\end{equation}
where $A$ is the area of a grid cell. When constructing DEM curves
$\Delta T$ is often not constant, but instead constant logarithmic
intervals are used. In this case we use temperature intervals of
$d\textrm{Log}_{10}T=0.1$. The intervals are furthermore over sampled in the
sense that intervals overlap in order to be able to construct
reasonably smooth curves.  

\ifnum\astroph=1
\FIGXIV
\fi

DEM curves for 5 selected points along loop 1, shown
as crosses in \Fig{fig:magneto70} (top), are shown in
\Fig{fig:DEMtr171}. It is very difficult to identify loop 1 from these
curves, and it is also clear that especially for 
points 2 and 3, the typical temperature visible in TRACE 171 is not
the temperature where most of the gas is. Point 5 shows especially
many isolated bumps, while point 3, even 
though it seems to be in the middle of a bright loop, show a clear
bump at low temperatures. These DEM curves are representative of most
of the ones we have produced and one can only seldomly find clear
signatures of loops, because of the ''pollution'' of loops seen in
images like \Fig{fig:magneto70} (top). In spite of these problems, it seems
that in this model, loops typically have a DEM of
$10^{21}-10^{22}\,\textrm{cm}^{-5}\,\textrm{K}^{-1}$ near 1 MK. This
level of emission measure is also what for instance
\citet{Schmelz+etal03} finds. In the same work \citep{Schmelz+etal03}
argue that because the maximum for their DEM curves along the loop is
not at the same temperature one has to be very careful in employing
this method. We agree on this point, but not on the conclusion
that the non constant maximum of the DEM curve is evidence for
a temperature gradient along the loop. The conclusion would hold true
if the loop were perfectly isolated in temperature. This is very
seldomly the case. By visual inspection, loop 1
seems to be reasonably isolated in \Fig{fig:magneto70} (top), but we
have seen that at the left leg this is not so, even though along
some of its length it is true. \Figure{fig:loop_point3} shows the
temperature of the gas and the 
contribution to the emission in TRACE 171 along the projected axis in
\Fig{fig:magneto70} (top) of point 4.
It is also clear that the sharp peak in emission is located in a
temperature dip, and that even though there are quite a few areas at a
temperature which should show up in the TRACE 171 filter, these are
not important, because the density is too low. The DEM is able to show
the range of temperatures involved but in spite of the sharp peak in
emission in the TRACE 171 filter, it is not clearly noticeable in the
DEM curve. It therefore seems that a simplistic approach to DEM curves
is a problematic way of trying to identify loops.

\ifnum\astroph=1
\FIGXV
\fi

The problem would become more severe when also the
resolution is worse than the typical width of a loop, since this
would introduce not only contamination along the line of sight, but
also in the 2D viewing plane. 

\section{Discussion and Conclusions}
The three loops discussed here are representative of a large number of loops we
have examined and illustrate that there are no static loops. Loops are
continuously changing, with 
large differences in heating, cooling and cross sections. It is
therefore futile to fit scaling relations to individual loops, since the
scaling properties will change with time and between loops, making any
scaling relations inexact. We have seen examples of loops in the
emulated TRACE pass bands that do not follow fieldlines, because the
field is sheared and heating is localised in thin structures across
fieldlines. The high thermal conductivity adjusts the temperature
gradient until conduction balances the heating rate, and makes
the effect hard to observe except when projection effects add up
contributions from the slightly brighter sheared heating
region. 

When looking at large
ensembles of loops in an active region the heating scales roughly with
the magnetic field strength squared and decreases exponentially from
the {\em{local}} height of the transition region. The scale height 
is essentially proportional to the mean separation of the main polarities in
the active region. 

Velocities in loops are assumed to be along fieldlines, and
that is what we find in the region just above the transition
region. Above roughly 10 Mm the velocities along the loops relative to
the total velocity amplitudes are
no longer close to unity, but span the whole interval between zero and
one. The velocity amplitudes are on average a few tens of km $\textrm{s}^{-1}$,
but velocities as high as 400 km $\textrm{s}^{-1}$ are present.

Density and pressure are less extreme in these loops than what has
been reported for long loops by \citet{Winebarger+etal03}, and are here only a
few times denser than the surroundings. It is, however, clear that the
density is approximately constant along loops. The loops therefore
show no decrease in emission measure with height. This simulation only
spans 37 Mm in height, slightly less than a pressure scale height for
a 1 million degree gas, and therefore we cannot say if higher loops
would show a decrease in density with height but since the heating
scale height also tends to increase with active region size,
we expect loops to be able to keep approximately constant density, or at least
have densities decreasing slower with height than for an atmosphere in
hydrostatic equilibrium.

Both heating and plasma flows are time dependent for loop 1, even
though it has an almost constant emission throughout the time span we
followed it. 
The coronal loops analysed here may be assumed to be generic,
but are most likely not representative of many of the loops
observed by TRACE and analysed in detail elsewhere since the latter
are biased towards being bright, isolated and 
long. The DEM curves measured for the bright TRACE 171 loop
do not easily reveal a loop, and as already concluded by
\citet{Schmelz+etal03}, differential emission measures must be 
treated carefully if they are to be used. 

The magnetic field in this work is only disturbed
slightly from a potential field, so large scale helicity is not
present.  Large scale helicity is often present in observed active regions
on the Sun and may be caused by large scale velocity fields, or may 
exist already in the flux 
emergence phase. Including disturbances able to create systematic
helicity would introduce unknown or arbitrary variables in contrast to the
ab initio approach of this work, but we do expect
that if a number of short loops in quiescent active regions are observed
in the TRACE or EIT 171 band similar loops as modelled here would be 
present.

Analysis of this simulation indicates that often-made assumptions, such
as those of hydrostatic equilibrium, a time independent and symmetric
heating function, and constant cross sections along loops are highly
questionable and cannot provide a basis for simplified models of loops.  
 
\acknowledgements{
BVG acknowledges support through an EC-TMR grant to the European
Solar Magnetometry Network.  The work of {\AA}N was supported in
part by the Danish Research Foundation, through its establishment
of the Theoretical Astrophysics Center. Computing time was provided 
by the Swedish
National Allocations Committee and by the Danish Center for Scientific
Computing.
}

\bibliographystyle{aa}
\bibliography{ms}
\ifnum\astroph=0
\clearpage
\FIGI
\clearpage
\FIGII
\clearpage
\FIGIII
\clearpage
\FIGIV
\clearpage
\FIGV
\clearpage
\FIGVI
\clearpage
\FIGVII
\clearpage
\FIGVIII
\clearpage
\FIGIX
\clearpage
\FIGX
\clearpage
\FIGXI
\clearpage
\FIGXII
\fi
\end{document}